\documentclass{SCIS2025}

\begin{document}
\ArticleType{RESEARCH PAPER}
\Year{2025}
\Month{January}
\Vol{68}
\No{1}
\DOI{}
\ArtNo{}
\ReceiveDate{}
\ReviseDate{}
\AcceptDate{}
\OnlineDate{}
\AuthorMark{}
\AuthorCitation{}

\title{Enhanced channel estimation for near-field IRS-aided multi-user MIMO system via a large deep residual network}{Title for citation}

\author[1]{Yan WANG}{}
\author[1]{Yongqiang LI}{}
\author[1]{Minghao CHEN}{}
\author[1]{Yu YAO}{}
\author[1,2,3]{Feng SHU}{{shufeng0101@163.com}}
\author[4]{\\Jiangzhou WANG}{}


\address[1]{School of Information and Communication Engineering, Hainan University, Haikou {\rm 570228}, China}
\address[2]{School of Information Science and Technology, Tibet University, Lhasa, {\rm 850000}, China}
\address[3]{School of Electronic and Optical Engineering, Nanjing University of Science and Technology, Nanjing {\rm 210094}, China}
\address[4]{School of Engineering, University of Kent, Canterbury CT2 7NT, U.K.}

\abstract{In this paper, the channel estimation (CE) of intelligent reflecting surface-aided near-field (NF) multi-user communication is investigated. Initially, the least square (LS) estimator and minimum mean square error (MMSE) estimator for the estimated channel are designed, their mean square errors (MSEs) are derived, and the Cram$\acute{e}$r-Rao lower bound (CRLB) is derived to serve as a benchmark for performance evaluation. 
Subsequently, in view of the fact that the NF channel model is more sensitive to distance variations compared to the far-field model, this leads to pronounced discrepancies in the user channel characteristics in different regions. 
To effectively capture and utilize these diverse channel features, users are initially divided into distinct regions predicated on pivotal parameters, such as channel angle and distance.
Correspondingly, a user region classifier based on convolutional neural networks is designed.
Then, to fully harness the potential of deep residual networks (DRNs) in denoising, the aforementioned CE problem is reconceptualized as a denoising task, and a DRN-driven single region NF CE network, named SR-DRN-NFCE, is proposed. 
In addition, by integrating SR-DRN-NFCE networks corresponding to different regions and conducting joint training in a federated learning (FL) manner, a new network is obtained, named FL-DRN-NFCE.
Simulation results demonstrate that the proposed FL-DRN-NFCE network outperforms LS, MMSE, and no residual connections in terms of MSE, and the proposed FL-DRN-NFCE method has higher CE accuracy than the SR-DRN-NFCE method.}


\maketitle

\section{Introduction}

Intelligent reflecting surfaces (IRSs), classified as either passive \cite{Shu2021Enhanced} or active \cite{Wang2023Asymptotic}, can accurately transmit a signal, thereby improving the coverage and communication efficiency of that signal \cite{Yu2021Low-cost}. 
However, the realization of this advantage relies on the precise acquisition of channel state information (CSI) \cite{Hu2020Statistical}.
In IRS-assisted wireless scenarios, the introduction of IRSs increases the complexity and uncertainty of signal propagation paths \cite{Tao2023Joint}, making traditional channel estimation (CE) methods difficult to apply directly.
Consequently, how to efficiently and accurately estimate the complete CSI, including the influence of IRS, has become an urgent and challenging core task in the research field of IRS-aided wireless technology \cite{Wang2022Beamforming}.

\subsection{Prior works}

A diversity of new algorithms and techniques have been investigated by scholars for the CE problem of IRS-aided communication networks \cite{Zheng2022A-Survey}. 
For example, three channel estimators were developed by Wang et al. \cite{Wang2023One-Bit} for the uplink CE of IRS-assisted multi-user (MU) massive multiple-input single-output (MISO) systems.
Specifically, the first estimator cleverly utilized the element sparsity of cascaded channels for optimization; the second estimator delved into the sparsity of the common row structure inherent in the IRS to base station (BS) channel; finally, the third estimator further integrated the sparsity of common row and column structures induced by the limited scattering environment around the user, thereby achieving more refined CE.
In addition, a Bayesian CE method for IRS-enabled millimeter wave (mmWave) massive multiple-input multiple-output (MIMO) networks was proposed by Kim et al. \cite{Kim2023Bayesian}.
Numerical analysis indicated that the introduction of active sensors significantly reduced the needed training cost of the designed CE algorithm.
Subsequently, a two-stage strategy was proposed by Chen et al. \cite{Chen2024Joint} for estimating the sensing and communication channels in a self-sensing IRS-aided mmWave integrated sensing and communication (ISAC) network.
Experimental results verified the availability of the constructed two-stage scheme.
In 2022, two novel anchor-assisted CE methods were investigated by Guang et al. \cite{Guan2022Anchor-Assisted}, and the results showed that scheme I was most suitable when the BS had a significant quantity of antennas, whereas scheme II was more effective otherwise.
In addition to focusing on researching high-performance and low-complexity CE algorithms, a productive transmission protocol for CE and beam tracking was developed by Zheng et al. \cite{Zheng2022Intelligent}, which independently processed satellites as well as users in a distributed way, greatly lessening the complexity of implementation.

In addition, the inherent transceiver hardware impairments (HWIs) inevitably introduce additional noise and distortion, causing the training pilot signal and received signal to deviate from their ideal states, which in turn substantially increases the complexity of CE.
The inaccuracy of CE results will further adversely affect the system performance \cite{Li2024Stacked}. 
Therefore, the impact of HWI must be fully accounted for in the design of CE algorithms.
To achieve this objective, it may be necessary to integrate HWI models into the estimation process or adopt compensation techniques to effectively mitigate their negative effects \cite{Li2024EE}. 
Concurrently, the presence of HWI issues has motivated academic research toward exploring more robust CE approaches.
For instance, Li et al. \cite{Li2023Achievable} theoretically formulated the ergodic sum-rate of the IRS-assisted non-orthogonal multiple access (NOMA) uplink in the face of CE errors and HWI. Specifically, the IRS phase shift was configured based on the statistical CSI, followed by the linear minimum mean square error (LMMSE) CE of the equivalent channel spanning from the user to the BS.

Technically, traditional CE methods mainly rely on signal processing techniques and statistical models, such as least square (LS) CE and minimum mean square error (MMSE) CE.
These methods can estimate the channel state to a certain extent, but they are often limited by the complexity of the model and the limitations in prediction accuracy.
In particular, in complex and changeable wireless environments, the prediction accuracy and adaptability of traditional methods are often difficult to comply with the realistic demand.

Recently, deep learning (DL), as one of the core branches in the domain of artificial intelligence, has made extraordinary breakthroughs in two key areas, namely, image recognition and natural language processing, by virtue of its excellent ability in automatic feature extraction and complex pattern recognition.
The application of DL in the area of CE is expected to break through the drawbacks of traditional approaches and enhance the correctness and adaptability of CE.
For instance, Ye et al. \cite{Ye2022Channel} elaborated a conditional generative adversarial network (cGAN) architecture that could effectively estimate the characteristics of cascaded channels exploiting the received signal as a conditional variable. 
The experimental analysis fully verified that the cGAN approach exhibited excellent robust performance in the context of IRS-assisted communication networks.
Subsequently, a graph neural network (GNN)-based CE algorithm was proposed by Ye et al. \cite{Ye2024GNN-Based}, which could maintain good performance with low pilot overhead.
Moreover, existing neural network-based approaches typically involve the manual design of network architectures through a trial-and-error process, demanding extensive domain knowledge and human resources. 
Therefore, an automated method for constructing a high-performance neural network architecture for CE was designed by Shi et al. \cite{Shi2024Automatic}, which provided the most accurate estimation of CSI over other benchmark CE methods.
Specifically, Chu et al. \cite{Chu2024Adaptive} explored an adaptive and robust estimator for time-varying mmWave channels, considering the use of more general arbitrarily shaped IRS, in contrast to classical regular IRS elements arranged on a grid.
To reduce the complexity, a DL-based solver using fixed-point iteration and a cascaded DL network framework was studied.
The proposed DL-based method exhibits lower complexity when compared to the alternating optimization (AO)-based approach, making it suitable for highly dynamic communication scenarios.
In addition, as a critical information hub bridging CSI acquisition and network parameter optimization, the transmission efficiency and reliability of CSI feedback play a pivotal role in determining the overall performance of communication systems. 
To address this, a decoupled representation learning-enabled neural network was proposed by Xu et al. \cite{Xu2024Disentangled}, which leveraged the uplink CSI characteristics to achieve selective compression feedback.

However, most of the aforementioned CE approaches are based on assumptions of far-field (FF) channel models \cite{Wei2022Channel}.
This is because traditional wireless communication networks, i.e., the first generation (1G) to the fifth generation (5G), rely heavily on the spectrum below 6 GHz or even below 3 GHz.
Due to wavelength limitations, these networks typically employ smaller antenna arrays.
The combination of low-dimensional antenna arrays and lower frequencies typically limits the range of wireless near-field (NF) communications to a few meters, or even a few centimeters \cite{Cui2022Channel}.
Therefore, the depiction of traditional wireless communication networks is commonly based on FF assumptions.
The 6G network architectures are expected to introduce larger antenna aperture designs and incorporate high frequency band resources, e.g. centimeter-wave, mmWave, and even terahertz, which will considerably enhance the NF effect characteristics of the network.
The fusion application of emerging manners such as IRS \cite{Wang2023Intelligent}, extra large-scale MIMO (XL-MIMO), and cell-free MIMO, predicts that NF scenarios will become increasingly prevalent in future wireless communication. Consequently, the traditional assumption based on FF plane waves will no longer be the dominant analytical framework, and new theoretical channel models need to be sought to adapt to this change.
As a result, more and more researchers have embarked on relevant studies focusing on NF communication scenarios.

Characterization and estimation algorithms for NF sparse channels have become the focus of extensive academic attention.
Specifically, Pisharody et al. \cite{Pisharody2024Near-Field} innovatively proposed two algorithms aimed at accurately estimating the NF uplink channel in XL-MIMO systems. These algorithms cleverly exploited the spatial non-stationary characteristics unique to XL-MIMO channels, although the implementation was accompanied by a relatively high computational complexity.
In a follow-up study, a low-complexity sequential angle-distance CE (SADCE) scheme for NF XL-MIMO networks furnished with uniform planar arrays (UPAs) was designed by Huang et al. \cite{Huang2024Low-Complexity}. Notably, the prominent energy leakage effect in NF XL-MIMO channels constituted one of the key factors affecting the CE performance.
To cope with the difficult problem of compressed CE in XL-MIMO models, Guo et al. \cite{Guo2023Compressed} innovatively proposed a triple parametric decomposition (TPD) structure. The simulation verification showed that this TPD skeleton considerably refined the system performance over the current state-of-the-art technology.
In particular, an efficient model-based DL algorithm was investigated by Zhang et al. \cite{Zhang2023Near-Field} to estimate NF wireless channels for XL-MIMO communications.
This algorithm constructed a sparse dictionary model based on spatial grids, which transformed the NF CE task in XL-MIMO systems into an optimization problem under the compressed sensing (CS) framework. Subsequently, the learning iterative shrinkage and thresholding algorithm (LISTA) was used to efficiently solve the CS problem.

However, the existing NF CE methods are mostly focused on XL-MIMO systems without IRS assistance, whereas corresponding CE methods for IRS-aided NF communication environments have also been explored and investigated in depth.
A low complexity CE strategy based on the assumption of NF spherical wavefront was designed by Yang et al. \cite{Yang2023Channel}. 
To decrease the pilot overhead and computational complexity, this strategy simplified the CE process into two stages. The first stage focused on the angle domain variable estimation between the BS and users employing the hybrid beamforming structure, whereas the second stage further estimated the cascaded angle domain and polar domain factors of XL-IRS.
However, in this NF channel model, only the line-of-sight (LoS) component was considered, while the non-line-of-sight (NLoS) component was ignored.

\subsection{Our contributions}

Motivated by the aforementioned discussions, deep residual network (DRN)-based CE algorithms will be investigated for the IRS-aided MU NF communication.
To the best of our knowledge, there is relatively limited research on the application of DRN for CE in IRS-assisted MU NF communication systems, especially considering both direct and cascaded channels containing LoS and NLoS components.
The core contributions of this paper are condensed into the following succinct summary.

\begin{enumerate}
   \item  To begin with, an IRS-assisted MU NF communication system is constructed. For the NF channel model, unlike the work by Yang et al. \cite{Yang2023Channel}, which solely focused on the LoS component, the NLoS component is taken into account in this paper to offer a more holistic portrayal of the NF channel characteristics.
       Moreover, the CE problem for both direct and cascaded channels is concurrently investigated in this paper, in contrast to the prevalent focus solely on cascaded CE in most studies. Subsequently, the LS estimators for direct and cascaded channels are designed. Correspondingly, the closed-form expression for the mean square error (MSE) of the LS estimator is derived, and the Cram$\acute{e}$r-Rao lower bound (CRLB) is derived to serve as a benchmark for performance evaluation.
   \item  Considering the heightened sensitivity of NF channel models to distance variations compared to FF models, this leads to significant differences in channel characteristics among users across distinct regions. To capture and utilize these diversified channel characteristics, users are partitioned into distinct regions based on key parameters such as channel angle and distance. Accordingly, a user region classifier (RC) based on convolutional neural networks is designed. Simulation results show that the accuracy of the proposed RC approximates 95\% in the high signal-to-noise ratio (SNR) range.
   \item  The CE problem in IRS-assisted MU NF communication systems is re-envisioned as a denoising task, and a DRN-driven single region NF CE, namely, SR-DRN-NFCE network, is designed by leveraging the denoising advantages of DRN. Then, we propose an innovative CE framework that integrates SR-DRN-NFCE networks from different regions through collaborative training based on federated learning (FL), ultimately developing a novel network called FL-DRN-NFCE. Simulation results reveal that the proposed FL-DRN-NFCE network exhibits a lower level of MSE over the scheme without residual connections.
       In addition, compared with traditional methods such as LS and MMSE, the proposed FL-DRN-NFCE algorithm successfully reduces pilot overhead by five-sixths through the powerful feature extraction capabilities of DL technology.
\end{enumerate}

\subsection{Organization and notation}

The remainder of this paper is structured as follows. 
In Section 2, the system model of an IRS-aided MU NF communication is established. The CRLB is derived in Section 3. In Section 4, the RC, and the FL-DRN-NFCE network are proposed, and the computational complexity of the aforementioned networks is also given. Finally, the simulation results and conclusions are presented in Sections 5 and 6, respectively.

Notations: Throughout this paper, matrices are denoted by boldface uppercase letter (e.g., $\mathbf{A}$), vectors by boldface lowercase letter (e.g., $\mathbf{a}$), and scalars by lowercase letter, (e.g., $a$).
Moreover, $\mathbb{C}^{a\times b}$ stands for the space of $a\times b$ matrices with complex entries. The real part is represented by $\mathfrak{Re}\{\cdot\}$, whereas the imaginary part is represented by $\mathfrak{Im}\{\cdot\}$. In addition, the sign $(\cdot)^{T}$ denotes the transpose operation, whereas $(\cdot)^{H}$ stands for the conjugate and transpose operation. 
The signs $\|\cdot\|_2$ and $\|\cdot\|_F$ stand for the 2-norm and $F$-norm, respectively. 
The notation $\mathbb{E}\{\cdot\}$ denotes the expectation operation, $\hat{[\;]}$ represents the estimation operation, 
and $\mathrm{tr}\{\cdot\}$ is the trace operation.
Furthermore, the identity matrix is denoted by $\mathbf{I}$, whereas the Kronecker and Hadamard products are respectively denoted by the symbols $\otimes$ and $\odot$.
In addition, $\text{diag}\{\mathbf{a}\}$ yields a diagonal matrix, where the primary diagonal elements are passively set to the components of $\mathbf{a}$.
Finally, $\lfloor\cdot\rfloor$ indicates rounding down.

\section{System model}
\begin{figure*}[h]
\centering
\includegraphics [width=0.85\textwidth]
{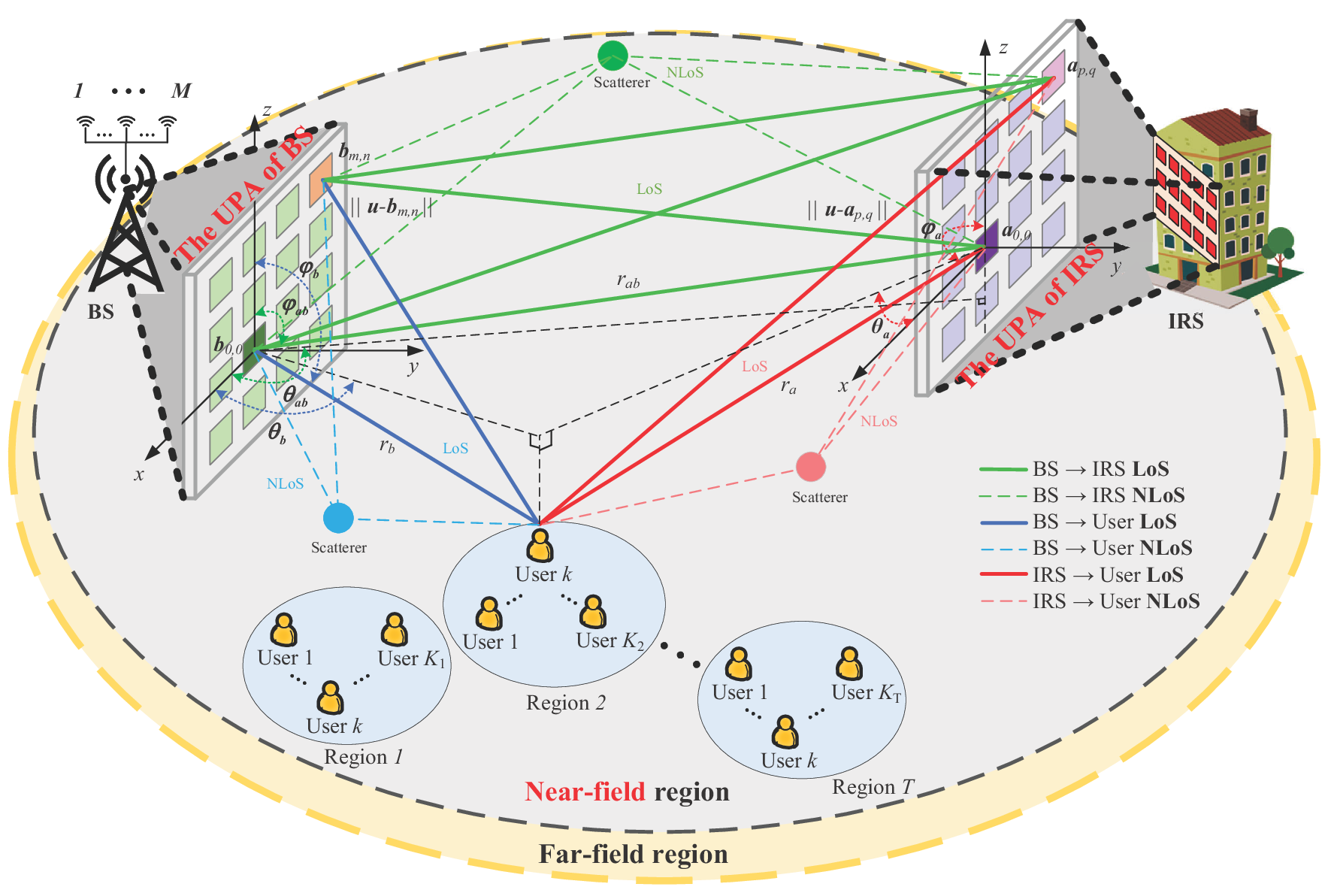}
\caption{The system model of an IRS-aided MU NF communication.}
\label{MUCE-system-model}
\end{figure*}

As sketched in Fig.~\ref{MUCE-system-model}, 
a MU NF wireless communication system, aided by an IRS, is taken into account, with IRS being strategically positioned between the BS and single-antenna users for communication augmentation.
The BS is furnished with an $M$-antenna array, while the IRS is composed of $N$ reflecting elements, both of which are set up in a UPA configuration.
In MU communication systems, as described in \cite{Dai2022Distributed}, 
a single user is often limited to information about a particular channel scenario, which leads to the possibility that neural network models trained based on the dataset collected by a single user may fail when the user crosses different regions.
In order to overcome this limitation and ensure that the neural network model performs well under a wide range of channel conditions, the whole cell is elaborately divided into $T$ independent regions, where $t = 1, 2, \ldots, T$.
For the convenience of subsequent analysis, the count of users in the $t$-th region is explicitly defined as $K_t$.
In particular, users in the same domain demonstrate highly similar channel characteristics, on the contrary, users distributed in distinct regions exhibit differentiated channel characteristics. 
Most importantly, when users migrate between regions, their channel characteristics may undergo notable changes, which puts forward higher demands for the design of neural network-based CE algorithms with high robustness.

Let $\mathbf{G}\in \mathbb{C}^{N\times M}$, $\mathbf{f}_{t, k}\in \mathbb{C}^{1\times N}$, and $\mathbf{h}_{t, k}\in \mathbb{C}^{1\times M}$ stand for the channels from BS to IRS, IRS to the $k$-th user in the $t$-th region, and BS to the $k$-th user in the $t$-th region, respectively, where $k=1, 2, \ldots, K_{t}$.
The received signal at the $k$-th user positioned in the $t$-th region can be modeled as follows
\begin{align}\label{ytk}
y_{t, k}=\mathbf{h}_{t,k}\mathbf{v}x+\mathbf{f}_{t,k}\mathbf{\Theta}\mathbf{Gv}x
+w_{t, k},
\end{align}
where $x$ represents the transmitted signal originating from the BS, and $\mathbf{v}\in \mathbb{C}^{M \times 1}$ denotes the transmit beamforming vector at BS. In addition, $\mathbf{\Theta}=\text{diag}\{\tilde{\boldsymbol{\theta}}\}\in\mathbb{C}^{N\times N}$ stands for the reflection coefficient matrix of the IRS, and $\tilde{\boldsymbol{\theta}}\in \mathbb{C}^{1\times N}$. Moreover, $w_{t, k}$ is the additive white Gaussian noise (AWGN) at the corresponding user with $w_{t, k}\sim \mathcal{CN}(0,\sigma_u^2)$.

Since $\mathbf{f}_{t,k}\mathbf{\Theta}
=\tilde{\boldsymbol{\theta}}
\text{diag}\{\mathbf{f}_{t,k}\}$, it can be intuitively deduced that $\mathbf{f}_{t,k}\mathbf{\Theta}\mathbf{G}=
\tilde{\boldsymbol{\theta}}
\text{diag}\{\mathbf{f}_{t,k}\}
\mathbf{G}$.
Consequently, the core task of CE is to accurately estimate the following
\begin{align}
\mathbf{H}_{t,k}=
\begin{bmatrix}\mathbf{h}_{t,k}\\
\text{diag}\{\mathbf{f}_{t,k}\}
\mathbf{G}
\end{bmatrix}
\in \mathbb{C}^{(N+1)\times M}, \quad \forall~ t,k.
\end{align}
Correspondingly, (\ref{ytk}) can be transformed into another form of expression as follows
\begin{align}\label{ytktheta1}
y_{t, k}=
\boldsymbol{\theta}\mathbf{H}_{t,k}\mathbf{v}x
+w_{t, k},
\end{align}
where
\begin{align}
\boldsymbol{\theta}=[1\quad \tilde{\boldsymbol{\theta}}]\in \mathbb{C}^{1\times (N+1)}.
\end{align}

\subsection{Channel models}

\subsubsection{Near-field channel model}

Drawing inspiration from \cite{Liu2023Near-field-Open-J}, it is assumed that the $M$ antennas of BS's UPA are arranged in an array of $M_{xb} \times M_{zb}$ on the $x-z$ plane, namely $M = M_{xb} \times M_{zb}$, where $M_{xb} = 2 \tilde{M}_{xb}+1$ and $M_{zb} = 2 \tilde{M}_{zb} +1$, and the antenna spacing of this array is set to $d_{xb}$ and $d_{zb}$ in the two dimensions, respectively.
Additionally, $\theta_b$ and $\phi_b$ signify the azimuth angle and elevation angle of the user relative to the $x-z$ plane of BS, respectively.
Also, the distance between the user and the central of BS's UPA can be represented as $r_{b}$. 
As a result, the coordinates of the $k$-th user situated in the $t$-th region can be given as
\begin{align}
\mathbf{u}_b
=(r_{b}\cos\theta_{b}\sin\phi_{b},~
r_{b}\sin\theta_{b}\sin\phi_{b},~
r_{b}\cos\phi_{b}).
\end{align}
Similarly, the coordinates of the $(m, n)$-th antenna of the BS's UPA are obtained as follows
\begin{align}
\mathbf{b}_{m,n}=(nd_{xb},~0,~md_{zb}),
\end{align}
where $\forall n\in\{-\tilde{M}_{xb},\ldots,\tilde{M}_{xb}\}$, and $\forall m\in\{-\tilde{M}_{zb},\ldots,\tilde{M}_{zb}\}$.

According to \cite{Liu2023Near-field-Open-J}, assuming $d_{xb}/r_b\ll1$ and $d_{zb}/r_b\ll1$, we can derive the propagation distance $\|\mathbf{u}_b-\mathbf{b}_{m,n}\|$ as follows
\begin{align}
\label{ubbmn}
\|\mathbf{u}_b-\mathbf{b}_{m,n}\|
&=
\sqrt{(r_b\cos\theta_b\sin\phi_b-nd_{xb})^2
+(r_b\sin\theta_b\sin\phi_b)^2
+(r_b\cos\phi_b-md_{zb})^2}\nonumber\\
&=r_b\sqrt{1+\frac{n^2d^2_{xb}}{r_b^2}
+\frac{m^2d^2_{zb}}{r_b^2}
-\frac{2r_bnd_{xb}\cos\theta_b\sin\phi_b}{r_b^2}
-\frac{2r_bmd_{zb}\cos\phi_b}{r_b^2}}.
\end{align}
Subsequently, by applying the following second-order Taylor expansion to (\ref{ubbmn})
\begin{align}\label{SOTaylor}
\sqrt{1+x}\approx1+\frac12x
-\frac18x^2+\mathcal{O}(x^2),
\end{align}
then, (\ref{ubbmn}) can be approximated as
\begin{align}
\|\mathbf{u}_b-\mathbf{b}_{m,n}\|
\approx r_b\underbrace{-nd_{xb}\cos\theta_b\sin\phi_b
+\frac{n^2d_{xb}^2
{\left(1-\cos^2\theta_b\sin^2\phi_b\right)}}
{2r_b}}_{h_1,~\text{only depend on $n$}}
\underbrace{-md_{zb}\cos\phi_b
+\frac{m^2d_{zb}^2\sin^2\phi_b}{2r_b}}
_{h_2,~\text{only depend on $m$}},
\end{align}
where $h_1$ and $h_2$ solely contingent on $n$ and $m$, respectively.

Referring to the uniform spherical wave (USW) model of \cite{Starer1994Passive}, \cite{Bjornson2021A}, the NF LoS MISO channel can be formulated as follows
\begin{align}
\mathbf{h}_{t, k}^\mathrm{LoS}
=\beta_b e^{-j\frac{2\pi}{\lambda}r_b}
\bigg[e^{-j\frac{2\pi}{\lambda}
\left(\big\|\mathbf{u}_b
-\mathbf{b}_{-\tilde{M}_{xb}, -\tilde{M}_{zb}}\big\|
-r_b\right)},\ldots,
e^{-j\frac{2\pi}{\lambda}
\left(\big\|\mathbf{u}_b
-\mathbf{b}_{\tilde{M}_{xb}, \tilde{M}_{zb}}\big\|
-r_b\right)}\bigg]^{T}.
\end{align}
Once the constant phase $e^{-j\frac{2\pi}{\lambda}r_b}$ is taken out, the resulting phase of the Array Response Vector (ARV) is obtained, referred to as $h_1$ and $h_2$.
More specifically, the ARV of BS-user channel is as follows
\begin{align}
\mathbf{a}(\theta_b,\phi_b,r_b)
=\mathbf{a}_{x}(\theta_b,\phi_b,r_b)
\otimes\mathbf{a}_{z}(\phi_b,r_b), 
\end{align}
where
\begin{align}
\left[\mathbf{a}_{x}
(\theta_b,\phi_b,r_b)\right]_{n}=
e^{-j\frac{2\pi}\lambda\cdot h_1}=
e^{-j\frac{2\pi}\lambda
\left(-nd_{xb}\cos\theta_b\sin\phi_b
+\frac{n^2d_{xb}^2
\left(1-\cos^2\theta_b\sin^2\phi_b\right)}
{2r_b}\right)},
\end{align}
and
\begin{align}
\left[\mathbf{a}_z(\phi_b,r_b)\right]_m
=
e^{-j\frac{2\pi}\lambda\cdot h_2}=
e^{-j\frac{2\pi}{\lambda}
\left(-md_{zb}\cos\phi_b
+\frac{m^2d_{zb}^2\sin^2\phi_b}{2r_b}\right)}.
\end{align}

As demonstrated in Fig.~\ref{MUCE-system-model}, the multipath effect in NF environment, originating from the refraction and reflection of signals triggered by scatterers, prompts the signals to reach the users via NLoS paths, and this stochasticity endows the channel with dynamic characteristics, which urgently requires a statistical model to capture its changing patterns.
In this context, the channel from BS to scatterer can be analogized to the MISO channel model, and thus, a novel expression of NF multipath channel can be constructed as follows
\begin{align}
\mathbf{h}_{t, k}
=\underbrace{\beta_b\mathbf{a}
(\theta_b, \phi_b, r_b)}_{\mathrm{LoS}}
+\sum_{\ell=1}^{L_b}
\underbrace{\tilde{\beta}_{b, \ell}\mathbf{a}
(\tilde{\theta}_{b, \ell}, \tilde{\phi}_{b, \ell}, \tilde{r}_{b, \ell})}_{\mathrm{NLoS}}.
\end{align}
where $L_b$ denotes the number of scatterers between the BS and the user, and $\ell=1,2,\ldots,L_b$. Moreover, $\tilde{\theta}_{b, \ell}$, $\tilde{\phi}_{b, \ell}$, $\tilde{r}_{b, \ell}$, and $\tilde{\beta}_{b, \ell}$ represent the corresponding azimuth, elevation, distance, and channel gain of the $\ell$-th scatterer, respectively.

Likewise, the UPA of IRS is also placed in the $x-z$ plane, parallel to BS's UPA.
For the $k$-th user positioned in the $t$-th region, their coordinates relative to IRS's UPA are articulated by 
\begin{align}
\mathbf{u}_a=(r_{a}\cos\theta_{a}\sin\phi_{a},
r_{a}\sin\theta_{a}\sin\phi_{a},
r_{a}\cos\phi_{a}),
\end{align}
where $\theta_{a}$, $\phi_{a}$, and $r_{a}$ are the user's azimuth angle and elevation angle with respect to the IRS, respectively, and the user's distance from the IRS center element.
Correspondingly, the $(p, q)$-th element of the IRS's element array are given by 
\begin{align}
\mathbf{a}_{p,q}=(qd_{xa},~0,~pd_{za}),~\forall q\in\{-\tilde{N}_{xa},\ldots,\tilde{N}_{xa}\},
~p\in\{-\tilde{N}_{za},\ldots,\tilde{N}_{za}\},
\end{align}
where $N = N_{xa} \times N_{za}$, and $N_{xa} = 2 \tilde{N}_{xa}+1$ and $N_{za} = 2 \tilde{N}_{za} +1$. Correspondingly, the element spacing of IRS's UPA along the $x$ and $z$ axis is represented by $d_ {xa} $and $d_ {za} $, respectively.

Similar to (\ref{ubbmn}), by employing (\ref{SOTaylor}), the distance between the $k$-th user in the $t$-th region and the $(p, q)$-th element of the IRS is as follows
\begin{align}
\|\mathbf{u}_a-\mathbf{a}_{p,q}\|
\approx r_a
\underbrace{-qd_{xa}\cos\theta_a\sin\phi_a
+\frac{q^2d_{xa}^2
{\left(1-\cos^2\theta_a\sin^2\phi_a\right)}}
{2r_a}}_{f_1,~\text{only depend on $q$}}
\underbrace{-pd_{za}\cos\phi_a
+\frac{p^2d_{za}^2\sin^2\phi_a}{2r_a}}
_{f_2,~\text{only depend on $p$}},
\end{align}
thus, referring to \cite{Liu2023Near-field-Open-J}, the ARV of IRS-user channel can be displayed as follows
\begin{align}
&\mathbf{a}(\theta_a,\phi_a,r_a)
=\mathbf{a}_{x}(\theta_a,\phi_a,r_a)
\otimes\mathbf{a}_{z}(\phi_a,r_a), \end{align}
where
\begin{align}
\left[\mathbf{a}_{x}(\theta_a,\phi_a,r_a)\right]_{q}
=
e^{-j\frac{2\pi}\lambda\cdot f_1}
=e^{-j\frac{2\pi}\lambda
\left(-qd_{xa}\cos\theta_a\sin\phi_a
+\frac{q^2d_{xa}^2
\left(1-\cos^2\theta_a\sin^2\phi_a\right)}
{2r_a}\right)},
\end{align}
and
\begin{align}
\left[\mathbf{a}_z(\phi_a,r_a)\right]_p
=
e^{-j\frac{2\pi}\lambda\cdot f_2}
=e^{-j\frac{2\pi}{\lambda}
\left(-pd_{za}\cos\phi_a
+\frac{p^2d_{za}^2\sin^2\phi_a}{2r_a}\right)}.
\end{align}
As depicted in Fig.~\ref{MUCE-system-model}, the presence of scatterers within the environment can elicit multipath effects in the IRS-user NF channel. Hence, the channel between the $k$-th user in the $t$-th region and the IRS can be formulated as
\begin{align}
\mathbf{f}_{t, k}
=\underbrace{\beta_a\mathbf{a}
(\theta_a, \phi_a, r_a)}_{\mathrm{LoS}}
+\sum_{\ell=1}^{L_a}
\underbrace{\tilde{\beta}_{a, \ell}\mathbf{a}
(\tilde{\theta}_{a, \ell}, \tilde{\phi}_{a, \ell}, \tilde{r}_{a, \ell})}_{\mathrm{NLoS}},
\end{align}
where $L_a$ is the count of scatterers between IRS and user, and the azimuth angle, elevation angle, and distance of the $\ell$-th scatterer are represented as $\tilde{\theta}_{a, \ell}$, $\tilde{\phi}_{a, \ell}$, and $\tilde{r}_{a, \ell}$. 

As can be illustrated in Fig.~\ref{MUCE-system-model}, the NLoS MIMO path can be conceptualized as the fusion of two distinct MISO channels with respect to the BS and IRS, respectively. 
As a result, the channel between BS and IRS can be written as follows
\begin{align}
\mathbf{G}=\mathbf{G}_{\mathrm{LoS}}
+\sum_{\ell=1}^{L_{ba}}
\underbrace{\tilde{\beta}_{\ell}\mathbf{a}
(\tilde{\theta}_{ba, \ell}, \tilde{\phi}_{ba, \ell}, \tilde{r}_{ba, \ell})\mathbf{a}^T
(\tilde{\theta}_{ab, \ell}, \tilde{\phi}_{ab, \ell}, \tilde{r}_{ab, \ell})}_{\mathrm{NLoS}},
\end{align}
where the quantity of scatterers is designated as $L_{ba}$. In addition, $\tilde{\theta}_{ba, \ell}$ ($\tilde{\theta}_{ab, \ell}$) is the azimuth angle of the $\ell$-th scatterer with respect to the $x-z$ plane of IRS (BS), and the elevation angle of the $\ell$-th scatterer with respect to the $x-z$ plane of IRS (BS) is signified as $\tilde{\phi}_{ba, \ell}$ ($\tilde{\phi}_{ab, \ell}$). In particular, $\tilde{r}_{ba, \ell}$ ($\tilde{r}_{ab, \ell}$) denotes the distance between the $\ell$-th scatterer and the central element of IRS (BS). More precisely, according to \cite{Liu2023Near-field-Open-J}, $\mathbf{G}_{\mathrm{LoS}}$ can be embodied as follows
\begin{align}\label{Gchannel}
\mathbf{G}_{\mathrm{LoS}}
=\tilde{\beta}\mathbf{a}
(\theta_{ba},\phi_{ba},r_{ba})\Big(\mathbf{a}
(\theta_{ab},\phi_{ab},r_{ab})\Big)^{T}
\odot
\big(\mathbf{G}_c^x\otimes\mathbf{G}_c^z\big),
\end{align}
where
\begin{align}\label{Gchannelx}
\left[\mathbf{G}_c^x\right]_{q,n}
=e^{-j\frac{2\pi}
{\lambda r_{ab}}nd_{xb}qd_{xa}
\left(1-\cos^2\theta_{ab}\sin^2\phi_{ab}\right)},
\end{align}
and
\begin{align}\label{Gchannelz}
\left[\mathbf{G}_c^z\right]_{p,m}
=e^{-j\frac{2\pi}{\lambda r_{ab}}md_{zb}pd_{za}\sin^2\phi_{ab}}.
\end{align}
In (\ref{Gchannel}), (\ref{Gchannelx}), and (\ref{Gchannelz}), $\theta_{ba}$ ($\theta_{ab}$) and $\phi_{ba}$ ($\phi_{ab}$) represent the azimuth and elevation angles of the center of BS (IRS) relative to the center of IRS (BS), respectively.
Besides, $r_{ba}=r_{ab}
=\|\mathbf{a}_{0,0}-\mathbf{b}_{0,0}\|$ denote the distance between the central of IRS and the central of BS.
As evident from (\ref{Gchannel}), the NF LoS MIMO path matrix between parallel UPAs incorporates a interleaved variable, namely $\mathbf{G}_c^x\otimes\mathbf{G}_c^z$, which cannot be simply decomposed into the direct product of the ARVs of BS-side and IRS-side.

\subsubsection{Far-field channel model}

As a comparison, the FF ARV can be specifically described as follows
\begin{align}
\mathbf{a}_{\mathrm{far}}{(\theta_b,\phi_b)}
=&\left[e^{-j\frac{2\pi}\lambda
\tilde{M}_{xb}d_{xb}\cos\theta_b\sin\phi_b},
\ldots,
e^{j\frac{2\pi}\lambda\tilde{M}_{xb}d_{xb}
\cos\theta_b\sin\phi_b}\right]^{T}\nonumber\\
&\otimes
\left[e^{-j\frac{2\pi}\lambda\tilde{M}_{zb}d_{zb}
\cos\phi_b},\ldots,e^{j\frac{2\pi}\lambda
\tilde{M}_{zb}d_{zb}\cos\phi_b}\right]^{T}.
\end{align}
Correspondingly, in this case, according to \cite{Liu2023Near-field-Open-J}, the FF channel model between BS-user, IRS-user, and BS-IRS can be modeled as follows
\begin{align}
\mathbf{h}_{t, k}^{\text{far}}
=\underbrace{\beta_b\mathbf{a}_{\mathrm{far}}
(\theta_b, \phi_b)}_{\mathrm{LoS}}
+\sum_{\ell=1}^{L_b}
\underbrace{\tilde{\beta}_{b, \ell}
\mathbf{a}_{\mathrm{far}}
(\tilde{\theta}_{b, \ell}, \tilde{\phi}_{b, \ell})}_{\mathrm{NLoS}},
\end{align}
\begin{align}
\mathbf{f}_{t, k}^{\text{far}}
=\underbrace{\beta_a\mathbf{a}_{\mathrm{far}}
(\theta_a, \phi_a)}_{\mathrm{LoS}}
+\sum_{\ell=1}^{L_a}
\underbrace{\tilde{\beta}_{a, \ell}
\mathbf{a}_{\mathrm{far}}
(\tilde{\theta}_{a, \ell}, \tilde{\phi}_{a, \ell})}_{\mathrm{NLoS}},
\end{align}
and
\begin{align}
\mathbf{G}^{\text{far}}
=\mathbf{G}_{\mathrm{LoS}}^{\text{far}}
+\sum_{\ell=1}^{L_{ba}}
\underbrace{\tilde{\beta}_{\ell}
\mathbf{a}_{\mathrm{far}}
(\tilde{\theta}_{ba, \ell}, \tilde{\phi}_{ba, \ell})
\mathbf{a}^T_{\mathrm{far}}
(\tilde{\theta}_{ab, \ell}, \tilde{\phi}_{ab, \ell})}_{\mathrm{NLoS}},
\end{align}
in which, in the FF scenario, the LoS component between the BS and IRS channels can be expressed as
\begin{align}
\mathbf{G}_{\mathrm{LoS}}^{\text{far}}
=\tilde{\beta}\mathbf{a}_{\mathrm{far}}
(\theta_{ba},\phi_{ba})
\Big(\mathbf{a}_{\mathrm{far}}
(\theta_{ab},\phi_{ab})\Big)^{T}.
\end{align}

\subsection{Problem formulation}

To ascertain the downlink direct and cascaded channel, i.e. $\mathbf{H}_{t,k}$, BS transmits predetermined pilot signals to users, facilitated by the IRS, spanning across $Q$ time slots, where $q=1, 2, \ldots, Q$.
Drawing upon (\ref{ytktheta1}), 
$y^{p}_{t, k, q}$ for the $k$-th user in the $t$-th region, in the $q$-th time slot, is mathematically formulated as follows
\begin{align}
y^{p}_{t, k, q}=
\boldsymbol{\theta}_q
\mathbf{H}_{t,k}\mathbf{v}p_{q}
+w_{t, k, q},
\end{align}
where the pilot signal transmitted by BS is signified as $p_{q}$, and $\boldsymbol{\theta}_q$ stands for the IRS phase shift within the current time slot.
Also, $w_{t, k, q}$ denotes the noise at the user of the current time slot.

Following the transmission of pilot signals across $Q$ consecutive time slots, we derive the comprehensive $Q \times 1$ received pilot vector as depicted below.
\begin{align}\label{ytkpilot}
\mathbf{y}^{p}_{t, k}
&=[y^{p}_{t, k, 1},~y^{p}_{t, k, 2},\ldots,~
y^{p}_{t, k, Q}]^T
=\boldsymbol{\Phi}\mathbf{H}_{t,k}
\mathbf{v}p_{t, k, q}
+\mathbf{w}_{t, k},
\end{align}
where $\boldsymbol{\Phi}=[\boldsymbol{\theta}_1^T,
~\boldsymbol{\theta}_2^T,\ldots,~
\boldsymbol{\theta}_Q^T]^T
\in\mathbb{C}^{Q\times(N+1)}$ and $\mathbf{w}_{t, k}
=[w_{t, k, 1},~w_{t, k, 2},\ldots,~w_{t, k, Q}]^T\in\mathbb{C}^{Q\times 1}$.
According to \cite{Dai2022Distributed}, assuming $p_{t, k, q}=1$ for the purpose of subsequent CE, (\ref{ytkpilot}) can be rephrased as
\begin{align}\label{p=1}
\mathbf{y}^{p}_{t, k}
=\boldsymbol{\Phi}\mathbf{H}_{t,k}
\mathbf{v}
+\mathbf{w}_{t, k}.
\end{align}
Due to the fact that $\mathrm{vec}(\mathbf{ABC}) = (\mathbf{C}^T \otimes \mathbf{A})\mathrm{vec}(\mathbf{B})$, (\ref{p=1}) can be reformulated as
\begin{align}\label{QH}
\mathbf{y}^{p}_{t, k}
&=\underbrace{(\mathbf{v}^T \otimes \boldsymbol{\Phi})
}_{\mathbf{A}_{t, k}}
\underbrace{
\mathrm{vec}(\mathbf{H}_{t,k})}
_{\mathbf{h}^{\prime}_{t,k}}
+\mathbf{w}_{t, k}
=\mathbf{A}_{t, k}\mathbf{h}^{\prime}_{t,k}
+\mathbf{w}_{t, k}.
\end{align}
It follows from (\ref{QH}) that the essence of CE lies in reconstructing $\mathbf{h}^{\prime}_{t,k}$ utilizing the known $\mathbf{y}^{p}_{t, k}$ and $\mathbf{A}_{t, k}$.
According to \cite{Biguesh2006Training-Based}, the estimated $\mathbf{h}^{\prime}_{t,k}$ using LS estimator is provided by
\begin{align}\label{35}
\widehat{\mathbf{h}}^{\prime}_{t,k,\mathrm{LS}}=
\mathbf{A}_{t, k}^{\dag}\mathbf{y}^{p}_{t, k},
\end{align}
where $\mathbf{A}_{t, k}^{\dag}=\mathbf{A}_{t, k}^{H}(\mathbf{A}_{t, k}\mathbf{A}_{t, k}^{H})^{-1}$ denotes the pseudoinverse of $\mathbf{A}_{t, k}$.

The estimation MSE of (\ref{35}) is
\begin{align}
\varepsilon_{\mathrm{LS}}&=
\mathbb{E}\{\|\mathbf{h}^{\prime}_{t,k}
-\widehat{\mathbf{h}}^{\prime}_{t,k,\mathrm{LS}}
\|_F^2\}
=\mathbb{E}\{\|\mathbf{h}^{\prime}_{t,k}
-\mathbf{A}_{t, k}^{\dag}(\mathbf{A}_{t, k}\mathbf{h}^{\prime}_{t,k}
+\mathbf{w}_{t, k})
\|_F^2\}
=\sigma_u^2
\mathrm{tr}\big(
(\mathbf{A}_{t, k}\mathbf{A}_{t, k}^{H})^{-H}\big).
\end{align}
It is worth noting that since $\mathbf{A}_{t, k}=\mathbf{v}^T \otimes \boldsymbol{\Phi}$, $\mathbf{v}$ and $\boldsymbol{\Phi}$ need to be pre-determined to be constant values as a prerequisite for CE.
Consequently, according to \cite{Rao2014Distributed}, the transmit beamforming vector $\mathbf{v}$ randomly selects their components from $\big\{-\frac{1}{\sqrt{M}}, +\frac{1}{\sqrt{M}}\big\}$. Furthermore, based on \cite{Liu2022Deep}, $\boldsymbol{\Phi}$ can be innovatively crafted as a discrete Fourier transform (DFT), namely,
\begin{align}
\boldsymbol{\Phi}^T=
\begin{bmatrix}1&1&\cdots&1
\\1&W_Q&\cdots&W_Q^{Q-1}
\\\cdots&\cdots&\ddots&\cdots
\\1&W_Q^N&\cdots&W_Q^{N(Q-1)}
\end{bmatrix}\in\mathbb{C}^{(N+1)\times Q},
\end{align}
where $W_Q=e^{j2\pi/Q}$.
It is worth noting that here, $p_{t, k, q}$,$\mathbf{v}$, $\boldsymbol{\Phi}$ are pre-designed to be fixed values for channel estimation according to \cite{Dai2022Distributed}, \cite{Rao2014Distributed}, \cite{Liu2022Deep}, respectively. However, related researchers have demonstrated that proper pilot optimization as well as IRS phase shift configuration will bring additional performance gains. Due to the page limit on paper, the optimization of pilot and IRS phase shift in channel estimation for IRS assisted communication systems will be our future research work.

Subsequently, the MMSE estimator for the channel $\mathbf{h}^{\prime}_{t,k}$ is
\begin{align}
\widehat{\mathbf{h}}^{\prime}_{t,k,\mathrm{MMSE}}=
\mathbf{W}_{\text{MMSE}}
\widehat{\mathbf{h}}^{\prime}_{t,k,\mathrm{LS}}.
\end{align}
Correspondingly, the MMSE estimator obtains an estimate based on $\mathbf{W}$ to minimize the MSE in the following given equation
\begin{align}\label{DirectvarepsilonMMSE}
\varepsilon_{\mathrm{MMSE}}=\mathbb{E}\{\|
\mathbf{h}^{\prime}_{t,k}
-\mathbf{W}_{\text{MMSE}}
\widehat{\mathbf{h}}^{\prime}_{t,k,\mathrm{LS}}\|_F^2\}.
\end{align}
By solving the problem presented in equation (\ref{DirectvarepsilonMMSE}), the following outcomes can be derived:
\begin{align}
\mathbf{W}_{\text{MMSE}}=
\mathbf{R}_{\mathbf{h}^{\prime}_{t,k}
\widehat{\mathbf{h}}^{\prime}_{t,k,\mathrm{LS}}}
\bigg(\mathbf{R}_{
\widehat{\mathbf{h}}^{\prime}_{t,k,\mathrm{LS}}
\widehat{\mathbf{h}}^{\prime}_{t,k,\mathrm{LS}}}
+\frac{\sigma_u^2}{\sigma_x^2}
\mathbf{I}\bigg)^{-1},
\end{align}
where $\mathbf{R}_{\mathbf{h}^{\prime}_{t,k}
\widehat{\mathbf{h}}^{\prime}_{t,k,\mathrm{LS}}}=
\mathbb{E}\{\mathbf{h}^{\prime}_{t,k}
\widehat{\mathbf{h}}^{\prime H}_{t,k,\mathrm{LS}}\}$ represents the cross-correlation matrix between the actual channel and the LS CE, while $\mathbf{R}_{
\widehat{\mathbf{h}}^{\prime}_{t,k,\mathrm{LS}}
\widehat{\mathbf{h}}^{\prime}_{t,k,\mathrm{LS}}}=
\mathbb{E}\{
\widehat{\mathbf{h}}^{\prime}_{t,k,\mathrm{LS}}
\widehat{\mathbf{h}}^{\prime H}_{t,k,\mathrm{LS}}\}$ denotes the autocorrelation matrix of the LS CE.

\section{CRLB}

The CRLB is often used to calculate the optimal estimation accuracy that can be achieved theoretically, which can evaluate the effectiveness of proposed CE algorithms.
For the CE problem (\ref{QH}), $\mathbf{A}_{t, k}$ is a real matrix, $\mathbf{y}^{p}_{t, k}\in\mathbb{C}^{Q\times 1}$ represents the known received signal, $\mathbf{h}^{\prime}_{t,k}$ denotes the channel to be estimated, and $\mathbf{w}_{t, k}$ is the whitened noise.
Consequently, (\ref{QH}) can be divided into two parts, where the real part is
\begin{align}\label{Re}
\mathbf{y}^{p}_{t, k,u}=\mathbf{A}_{t, k}\mathbf{h}^{\prime}_{t,k,u}+\mathbf{w}_{t, k,u},
\end{align}
and the imaginary part is as follows
\begin{align}\label{Im}
\mathbf{y}^{p}_{t, k,v}=\mathbf{A}_{t, k}\mathbf{h}^{\prime}_{t,k,v}+\mathbf{w}_{t, k,v},
\end{align}
where
\begin{align}
\mathbf{y}^{p}_{t, k,u}&=\mathfrak{Re}(\mathbf{y}^{p}_{t, k}),~
\mathbf{y}^{p}_{t, k,v}=\mathfrak{Im}(\mathbf{y}^{p}_{t, k}),\nonumber\\
\mathbf{h}^{\prime}_{t,k,u}
&=\mathfrak{Re}(\mathbf{h}^{\prime}_{t,k}),~
\mathbf{h}^{\prime}_{t,k,v}
=\mathfrak{Im}(\mathbf{h}^{\prime}_{t,k}),
\nonumber\\
\mathbf{w}_{t, k,u}&=\mathfrak{Re}(\mathbf{w}_{t, k}),~
\mathbf{w}_{t, k,v}=\mathfrak{Im}(\mathbf{w}_{t, k}).
\end{align}
Thus, the CRLB of $\hat{\mathbf{h}}^{\prime}_{t,k}$ can also be divided into two parts, which are shown as
\begin{align}
\gamma_{t,k}& =\gamma_{t,k}^{u}+\gamma_{t,k}^{v} 
=\mathbb{E}\left\{\left\|
\widehat{\mathbf{h}}^{\prime}_{t,k,u}
-\mathbf{h}^{\prime}_{t,k,u}\right\|^2\right\}
+\mathbb{E}\left\{\left\|
\widehat{\mathbf{h}}^{\prime}_{t,k,v}
-\mathbf{h}^{\prime}_{t,k,v}\right\|^2\right\}.
\end{align}
In this case, we first consider real part (\ref{Re}). Since the $\mathbf{w}_{t, k,u}$ follows the distribution of Gaussian distribution with 0 mean and $\sigma_u^2$ variance, the conditional probability density function of $\mathbf{y}^{p}_{t, k,u}$ with the given $\mathbf{h}^{\prime}_{t,k,u}$ is
\begin{align}
&p_{\mathbf{y}^{p}_{t, k,u}|\mathbf{h}^{\prime}_{t,k,u}}
\left(\mathbf{y}^{p}_{t, k,u};\mathbf{h}^{\prime}_{t,k,u}\right)
=\frac{1}{\left(2\pi\sigma_u^2\right)
^{Q/2}}
\exp\left\{-\frac{1}{2\sigma_u^2}\left\|
\mathbf{y}^{p}_{t, k,u}-\mathbf{A}_{t,k}\mathbf{h}^{\prime}_{t,k,u}\right\|^2\right\}.
\end{align}
The Fisher information matrix of (\ref{Re}) can then be derived as
\begin{align}
[\mathbf{J}]_{i,j}
&=-\mathbb{E}
\left\{\frac{p_{\mathbf{y}^{p}_{t, k,u}|\mathbf{h}^{\prime}_{t,k,u}}
\left(\mathbf{y}^{p}_{t, k,u};\mathbf{h}^{\prime}_{t,k,u}\right)}
{\partial h^{\prime}_{t,k,u,i}\partial h^{\prime}_{t,k,u,j}}\right\}
=\frac{1}{\sigma_u^2}
\left[\mathbf{A}_{t,k}^H\mathbf{A}_{t,k}\right]_{i,j},
\end{align}
where $h^{\prime}_{t,k,u,i}$, $h^{\prime}_{t,k,u,j}$ denote the $i$-th and $j$-th entry of $\mathbf{h}^{\prime}_{t,k,u}$. Then, the real part $\gamma_{t,k}^{u}$ is
\begin{align}
\gamma_{t,k}^{u}&=\mathbb{E}\left\{\left\|
\widehat{\mathbf{h}}^{\prime}_{t,k,u}
-\mathbf{h}^{\prime}_{t,k,u}\right\|^2\right\}
\geq\mathrm{tr}
\left\{\mathbf{J}_u^{-1}\right\}
=\sigma_u^2\mathrm{tr}
\left\{(\mathbf{A}_{t,k}^H
\mathbf{A}_{t,k})^{-1}\right\}.
\end{align}
Since $\mathbf{A}_{t,k}=\mathbf{v}^T \otimes \boldsymbol{\Phi}$, $(\mathbf{A}_{t,k}^H
\mathbf{A}_{t,k})^{-1}$ can be presented as
\begin{align}
(\mathbf{A}_{t,k}^H
\mathbf{A}_{t,k})^{-1}
&=\Bigg(\big(\mathbf{v}^T \otimes \boldsymbol{\Phi}\big)^H
\big(\mathbf{v}^T \otimes \boldsymbol{\Phi}\big)\Bigg)^{-1}
=\bigg(\big(\mathbf{v}\mathbf{v}^H\big)^{-1}\bigg)^T \otimes \big(\boldsymbol{\Phi}^H
\boldsymbol{\Phi}\big)^{-1}.
\end{align}
Since $\mathrm{tr}(\mathbf{A}
\otimes\mathbf{B})=\mathrm{tr}(\mathbf{A})
\mathrm{tr}(\mathbf{B})$ and $\mathrm{tr}(\mathbf{A}^T)=\mathrm{tr}(\mathbf{A})$, thus, $\mathrm{tr}
\left\{(\mathbf{A}_{t,k}^H
\mathbf{A}_{t,k})^{-1}\right\}$ can be calculated as
\begin{align}
\mathrm{tr}
\left\{(\mathbf{A}_{t,k}^H
\mathbf{A}_{t,k})^{-1}\right\}
&=\mathrm{tr}\Bigg(
\bigg(\big(\mathbf{v}\mathbf{v}^H\big)^{-1}
\bigg)^T \Bigg)
\mathrm{tr}\bigg(\big(
\boldsymbol{\Phi}^H
\boldsymbol{\Phi}\big)^{-1}\bigg)
=\mathrm{tr}
\bigg(\big(\mathbf{v}\mathbf{v}^H\big)^{-1}
\bigg)
\mathrm{tr}\bigg(\big(
\boldsymbol{\Phi}^H
\boldsymbol{\Phi}\big)^{-1}\bigg).
\end{align}
Assuming $\{\alpha_i\}_{i=1}^{M}$ is denoted as the $M$ eigenvalues of the matrix of $\mathbf{v}\mathbf{v}^H$, and $\{\beta_j\}_{j=1}^{N+1}$ is denoted as the $N+1$ eigenvalues of the matrix of $\boldsymbol{\Phi}^H
\boldsymbol{\Phi}$, then
\begin{align}
\mathrm{tr}
\left\{(\mathbf{A}_{t,k}^H
\mathbf{A}_{t,k})^{-1}\right\}
&=\Bigg(\sum_{i=1}^{M}\alpha_i^{-1}\Bigg)
\Bigg(\sum_{j=1}^{N+1}\beta_j^{-1}\Bigg).
\end{align}
According to \cite{Lu2023Near-field}, we have
\begin{align}
\mathrm{tr}
\left\{(\mathbf{A}_{t,k}^H
\mathbf{A}_{t,k})^{-1}\right\}
&
\geq M\Bigg(M/\sum_{i=1}^{M}\alpha_i\Bigg)
(N+1)\Bigg((N+1)/\sum_{j=1}^{N+1}\beta_j\Bigg)
=\frac{M^2}{\mathrm{tr}(\mathbf{v}\mathbf{v}^H)}
\frac{(N+1)^2}{\mathrm{tr}(\boldsymbol{\Phi}^H
\boldsymbol{\Phi})}.
\end{align}
Finally, the CRLB of the real part of $\hat{\mathbf{h}}^{\prime}_{t,k}$ becomes
\begin{align}
\gamma_{t,k}^{u}=\mathbb{E}\left\{\left\|
\widehat{\mathbf{h}}^{\prime}_{t,k,u}
-\mathbf{h}^{\prime}_{t,k,u}\right\|^2\right\}
=\sigma_u^2\frac{M(N+1)}{Q}.
\end{align}

Observing equations (\ref{Re}) and (\ref{Im}), it can be seen that the real part and imaginary part have the same form, namely
\begin{align}
\gamma_{t,k}^{v}=\gamma_{t,k}^{u}
=\sigma_u^2\frac{M(N+1)}{Q}.
\end{align}
Thus, the CRLB of (\ref{QH}) is as folows
\begin{align}
\gamma_{t,k}
=\gamma_{t,k}^{u}
+\gamma_{t,k}^{v}
=2\sigma_u^2\frac{M(N+1)}{Q}.
\end{align}

\section{Proposed G-DRN-NFCE algorithm}

\begin{figure*}[h]
\centering
\includegraphics [width=0.77\textwidth]{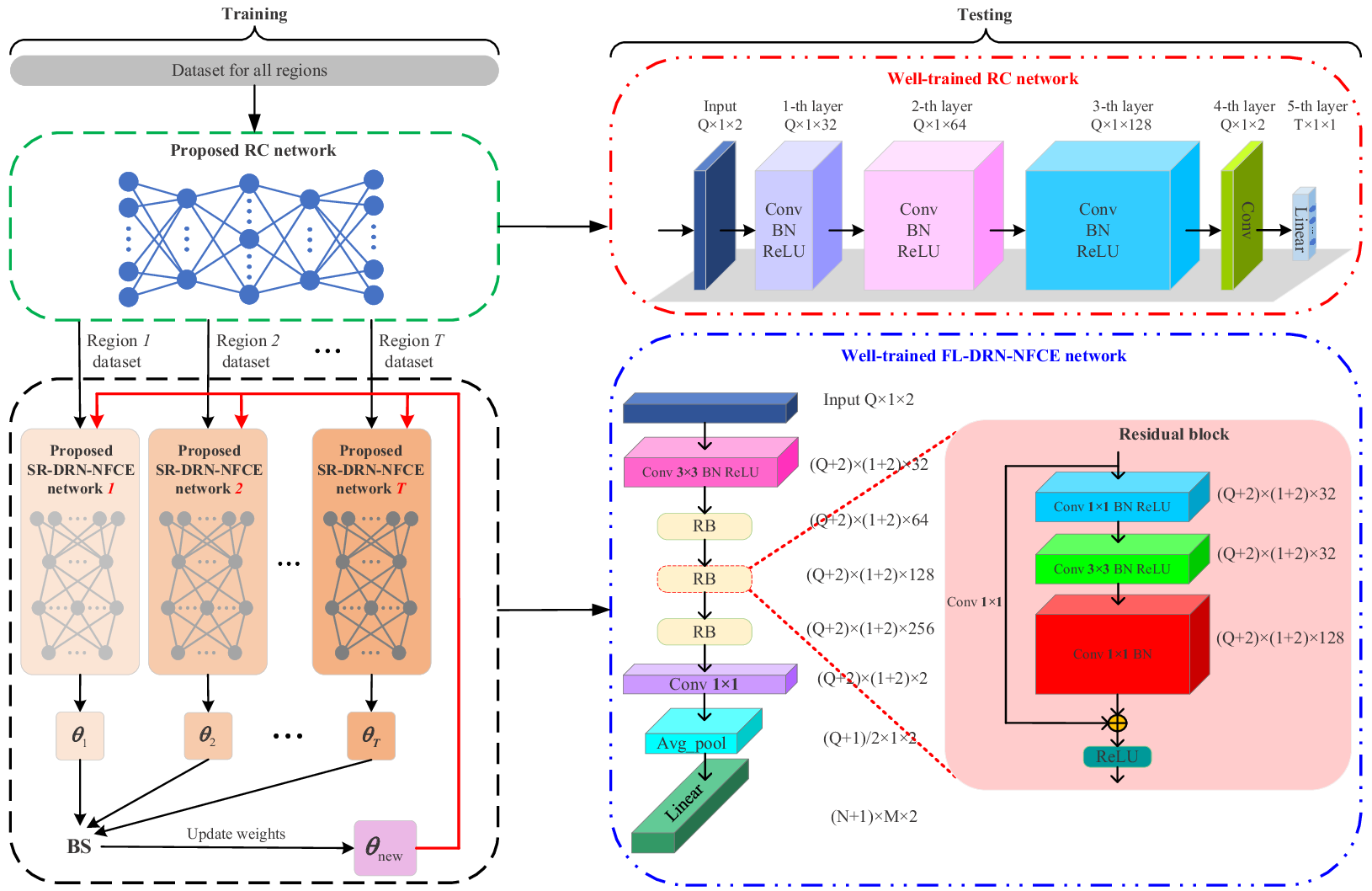}
\caption{The proposed G-DRN-NFCE framework.}
\label{DRN-NFCE-network}
\end{figure*}

In this section, as shown in Fig. \ref{DRN-NFCE-network}, the proposed global DRN-NFCE (G-DRN-NFCE) framework is introduced.
Firstly, the basic principles of the proposed SR-DRN-NFCE network are elaborated in detail in subsection 4.1.
Subsequently, in subsection 4.2, a CNN-based RC network is designed to precisely match datasets from different user regions with their corresponding SR-DRN-NFCE networks.
Then, the training details of the proposed FL-DRN-NFCE network are presented in subsection 4.3.
Finally, in subsection 4.4, the computational complexity of the above-mentioned networks is given.

\subsection{The basic principle of the proposed SR-DRN-NFCE network}

Specifically, the proposed SR-DRN-NFCE network introduces a non-linear function $\boldsymbol{f}_{\boldsymbol{\omega}}$ between $\mathbf{y}^{p}_{t, k}$ and $\widehat{\mathbf{h}}^{\prime}_{t,k,
\text{DRN}}$, which can be mathematically expressed as
\begin{align}
\widehat{\mathbf{h}}^{\prime}_{t,k,
\text{DRN}}&
=\boldsymbol{f}_{\boldsymbol{\omega}}
(\mathbf{y}^{p}_{t, k}),
\end{align}
where $\boldsymbol{\omega}$ represents weight. 
Accordingly, the expression for the loss function is as follows
\begin{align}
\mathcal{L}(\boldsymbol{\omega})
=\frac1{S_{t,k}}\sum_{s=1}^{S_{t,k}}\parallel
\widehat{\mathbf{h}}^{\prime}_{t,k,
\text{DRN}}
-\widehat{\mathbf{h}}^{\prime}_{t,k,\mathrm{LS}}
\parallel_2^2,
\end{align}
where $S_{t,k}$ stands for the size of training dataset. The traditional LS CE scheme can be used to obtain the label $\widehat{\mathbf{h}}^{\prime}_{t,k,\mathrm{LS}}$. The objective of exploiting the proposed SR-DRN-NFCE network for CE is to minimize $\mathcal{L}(\boldsymbol{\omega})$ by optimizing $\boldsymbol{\omega}$, which is 
\begin{align}
\min_{\boldsymbol{\omega}}\mathcal{L}
(\boldsymbol{\omega})
&=\frac{1}{S_{t,k}}\sum_{s=1}^{S_{t,k}}\parallel
\boldsymbol{f}_{\boldsymbol{\theta}}
(\mathbf{y}^{p}_{t, k})
-\widehat{\mathbf{h}}^{\prime}_{t,k,\mathrm{LS}}\parallel_{2}^{2}.
\end{align}
In each iteration $i$, $\boldsymbol{\omega}$ are updated through the following way, namely,
\begin{align}
\boldsymbol{\omega}_{i+1}
=\boldsymbol{\omega}_i
-\lambda_i\mathbf{g}
(\boldsymbol{\omega}_i),
\end{align}
where $\mathbf{g}(\boldsymbol{\omega}_i)$ is the gradient vector (GV) for $\boldsymbol{\omega}_i$, and $\lambda_i$ is the learning rate (LR).

\begin{table}[h]
\footnotesize
\caption{Hyperparameters of the proposed SR-DRN-NFCE network.}
\label{DRN}
\tabcolsep 14pt
\begin{tabular*}{\textwidth}{ccc}
\toprule
\multicolumn{3}{l}{\textbf{Input}}\\
\multicolumn{3}{l}{The received pilot signal $\mathbf{y}^{p}_{t, k}$ with the size of $Q \times 1$} \\\hline
\textbf{Layers} & \textbf{Operations} & \textbf{Number of parameters}  \\\hline
1 & Conv+BN+ReLU & $(3\times3\times2)\times 32$  \\
2 & RB  &$(1\times1\times32)\times 16+
(3\times3\times16)\times 16+
(1\times1\times16)\times 64+(1\times1\times32)\times 64$\\ 
3& RB  &$(1\times1\times64)\times 32+
(3\times3\times32)\times 32+
(1\times1\times32)\times 128+(1\times1\times64)\times 128$\\ 
4 & RB &$(1\times1\times128)\times 64+
(3\times3\times64)\times 64+
(1\times1\times64)\times 256+(1\times1\times128)\times 256$\\ 
5 & Conv  &$(1\times1\times256)\times 2$\\ 
6 & Avg-pool &-\\ 
7& Linear &$(\lfloor\frac{Q+1}{2}\rfloor\times1\times2)\times\big((N+1) \times M\times2\big)+(N+1) \times M\times2$\\\hline
\multicolumn{3}{l}{\textbf{Output}}\\
\multicolumn{3}{l}{The estimated channel with the size of $(N+1)M\times 1$} \\
\bottomrule
\end{tabular*}
\end{table}

Specifically, as displayed in Table~\ref{DRN}, the proposed SR-DRN-NFCE comprises two convolution (Conv) layers, one average pool (Avg-pool) layer, one linear layer, and three residual blocks (RBs).
The first layer is configured as a Conv layer with padding set to 2, a kernel size of 3, and a stride of 1. This design expands the input signal's dimensions in height and width from $Q\times1$ to $(Q+2)\times(1+2)$, aiming to meet the requirements of feature extraction in subsequent Conv layers and dimensionality reduction in Avg-pool layers.
Layers 2 to 4 consist of three consecutive RBs \cite{He2016Deepresidual} \cite{Sandler2018MobileNetV2}, characterized by their unique approach: initially reducing the dimensionality of the feature map using a $1\times1$ Conv kernel, followed by a standard Conv layer for feature extraction, and finally employing another $1\times1$ Conv kernel to expand the count of channels in the feature map. 
This design strategy offers significant advantages in reducing computational load and enhancing model efficiency. 
Additionally, to maintain dimensional consistency in the residual connections, an extra $1\times1$ Conv kernel is utilized for dimensionality adjustment.
The specific parameter configuration of the second RB is detailed in Fig.~\ref{DRN-NFCE-network}: the input feature map has dimensions of $(Q+2)\times(1+2)\times64$, and the output feature map has dimensions of $(Q+2)\times(1+2)\times128$. 
Within the 2nd RB, the number of channels is first reduced to 32, which halves the number of parameters in the intermediate Conv layer for feature extraction, thereby lowering computational complexity. 
The final $1\times1$ Conv layer merely changes the number of channels, with relatively low computational complexity.
The fifth layer is equipped with a $1\times1$ Conv layer that further reduces the count of channels in the input feature map to 2. 
The sixth layer is an Avg-pool layer with a kernel size of 3 and a stride of 2, which decreases the height and width of the input feature map from $(Q+2)\times(1+2)$ to $\lfloor(Q+1)/2\rfloor\times1$. 
The primary role of the fifth and sixth layers is to decrease the input dimensionality for the seventh linear layer, thereby effectively reducing the parameters and computational complexity of the linear layer.
Ultimately, the seventh linear layer is responsible for outputting the predicted signal, with an output dimension of $(N+1)\times M\times2$.

\subsection{The proposed RC network}

As diagrammed in Fig.~\ref{MUCE-system-model}, considering that the NF model is more sensitive to changes in distance, angle, and other factors than the FF model, which results in significant differences in the channel characteristics of users in different regions.
In order to effectively capture and fully utilize these diverse channel characteristics, the entire region is divided into $T$ different sub regions.
Therefore, in order to accurately match datasets from different user regions with corresponding SR-DRN-NFCE networks, a CNN-based RC is firstly designed.

\begin{table}[h]
\footnotesize
\caption{Hyperparameters of the proposed RC network.}
\label{region-classifier}
\tabcolsep 40pt
\begin{tabular*}{\textwidth}{ccc}
\toprule
\textbf{Layers} & \textbf{Operations} & \textbf{Number of parameters}  \\\hline
1 & Conv+BN+ReLU & $(3\times3\times2)\times 32$  \\
2 & Conv+BN+ReLU  &$(3\times3\times32)\times 64$\\ 
3& Conv+BN+ReLU  &$(3\times3\times64)\times 128$\\ 
4 & Conv &$(1\times1\times128)\times 2$\\  
5& Linear &$(Q\times1\times2)\times(3 \times 1\times1)+(3 \times 1\times1)$\\
\bottomrule
\end{tabular*}
\end{table}

As shown in Table \ref{region-classifier}, the proposed RC network architecture comprises four Conv layers and a linear layer. 
To enhance the steadiness of the model and expedite the training efficiency, a batch normalization (BN) is introduced between Conv and the rectified linear unit (ReLU).
The first three layers are all ``Conv + BN + ReLU'' operations, where the Conv layer is responsible for extracting features from the input signal. Importantly, the count of channels in these layers progressively increases from the first to the third. 
Specifically, the input of the proposed RC network is the received pilot signal $\mathbf{y}^{p}_{t, k}$, which includes both real and imaginary parts, therefore, the initial number of channels is 2.
Upon passing through the first Conv layer, the channel number increases to 32, enabling the network to capture a broader range of features. 
After the second Conv layer, the channel number escalates further to 64, enhancing the network's capacity to represent complex signal patterns. 
Finally, after the third Conv layer, the channel number reaches 128, significantly boosting the network's discriminative power and enabling it to better differentiate between similar signals, particularly in regions where signals may overlap or exhibit similar characteristics - this is crucial for capturing intricate patterns and subtle differences within the signal.
BN is employed to enhance training stability and accelerate convergence, while the ReLU activation function introduces nonlinearity, thereby bolstering the network's expressive capacity.
The fourth layer specifically utilizes a $1\times1$ Conv kernel for dimensionality reduction of the feature maps received from the preceding layer, reducing the channel number back to 2. 
This step effectively reduces the parameters and computational complexity of the subsequent linear layer.
The entire network is meticulously designed to efficiently extract signal features while mitigating computational load through dimensionality reduction. 
Ultimately, the linear layer maps the signals to category labels, achieving precise and efficient region classification of signals.

\subsection{The proposed FL-DRN-NFCE network}

Subsequently, as shown in Fig.~\ref{DRN-NFCE-network}, the user datasets of all regions are input into the well-trained RC network, then, users in each region calculate local GVs based on the current region's dataset.
As described in \cite{Xu2023Edge}\cite{Yao2024Wireless}, each user utilizes its local dataset to independently train and update local gradients, then uploads the trained local gradients to the BS to update the global model.
Once the BS has collected the local GVs of all regions, collaborative training is performed between the BS and the user in a FL manner, i.e., the BS updates the weights via the following manner
\begin{align}
\boldsymbol{\omega}_{i+1}
=\boldsymbol{\omega}_i
-\lambda_i\frac{1}{\sum\limits_{t=1}^TK_t}
\sum\limits_{t=1}^T
\sum\limits_{k=1}^{K_t}
\mathbf{g}_{t,k}(\boldsymbol{\omega}_i).
\end{align}
Finally, the well-trained FL-DRN-NFCE network is obtained in Fig.~\ref{DRN-NFCE-network}.
The well-trained FL-DRN-NFCE network is developed through a FL framework based on SR-DRN-NFCE networks corresponding to different regions. It maintains the same architecture configuration as the SR-DRN-NFCE network, but its weight parameters are specifically distinct.

To sum up, the step-by-step summary of the proposed G-DRN-NFCE algorithm is shown in \textbf{Algorithm 1}.

\begin{algorithm} 
\footnotesize
\renewcommand{\algorithmicrequire}
{\textbf{Training:}}	\renewcommand{\algorithmicensure}
{\textbf{Testing:}}
\caption{\textbf{Proposed G-DRN-NFCE algorithm}}
\label{FL-DRN-NFCE-algorithm}
\textbf{Initialization:} initialize trainable parameters, $i, j=0$, raw training data $\mathbf{y}^{p}_{t, k}$, $\mathbf{h}^{\prime}_{t,k}$
\begin{algorithmic}[1] 
\REQUIRE 
\STATE Input training dataset for all regions
\WHILE{$j<J$}
\STATE Update $\boldsymbol{\alpha}$ by backpropagation (BP) algorithm to minimize the loss function of the RC
\STATE $j=j+1$
\ENDWHILE
\STATE Output well-trained RC network 
\ENSURE 
\STATE Input testing set
\STATE \textbf{do} classification using proposed RC network in Fig.~\ref{DRN-NFCE-network}
\STATE Output classified dataset 
\REQUIRE 
\STATE Input training dataset for the current region 
\WHILE{$i<I_t$}
\STATE Update $\boldsymbol{\omega}$ by BP algorithm to minimize $\mathcal{L}(\boldsymbol{\omega})$
\STATE $i=i+1$
\ENDWHILE
\STATE Output well-trained SR-DRN-NFCE network 
\STATE Joint trainging of SR-DRN-NFCE networks corresponding to different regions in a FL manner
\ENSURE 
\STATE Input testing set
\STATE \textbf{do} channel estimation using proposed FL-DRN-NFCE network in Fig.~\ref{DRN-NFCE-network}
\STATE Output estimated channel 
\end{algorithmic}
\end{algorithm}

\subsection{Computational complexity analysis}

\subsubsection{For the proposed RC network}

The computational complexity of the 1st Conv \cite{Dong2019Deep} layer is
\begin{align}
C_{r1}=\mathcal{O}\Big(\big(3^2\cdot 2\cdot 32\big)\times\big(Q\cdot1\big)\Big)
=\mathcal{O}(576Q).
\end{align}
Similarly, the computational complexity of the 2rd and 3th Conv layers are 
\begin{align}
C_{r2}=\mathcal{O}\Big(\big(3^2\cdot 32\cdot 64\big)\times\big(Q\cdot1\big)\Big)
=\mathcal{O}(18432Q)
\end{align}
and
\begin{align}
C_{r3}=\mathcal{O}\Big(\big(3^2\cdot 64\cdot 128\big)\times\big(Q\cdot1\big)\Big)
=\mathcal{O}(73728Q),
\end{align}
respectively.
From Fig.~\ref{DRN-NFCE-network}, the computational complexity of the 4-th Conv layer is
\begin{align}
C_{r4}=\mathcal{O}\Big(\big(1\cdot 1\cdot 128\cdot2)\times\big(Q\cdot1\big)\Big)
=\mathcal{O}(256Q).
\end{align}
Under the premise of only considering multiplication operations, the computational complexity of the final linear layer \cite{Liu2023PoolNet} of the proposed RC network is as follows
\begin{align}
C_{r5}=\mathcal{O}\Big(\big(Q\cdot 1\cdot 2)\times\big(T\cdot1\cdot1\big)
\Big)
=\mathcal{O}(2QT).
\end{align}
Therefore, the total computational complexity of the proposed RC network is
\begin{align}
C_{r}&=C_{r1}+C_{r2}+C_{r3}+C_{r4}+C_{r5}
=\mathcal{O}(92992Q+2QT).
\end{align}

\subsubsection{For the proposed FL-DRN-NFCE network}

Based on Fig.~\ref{DRN-NFCE-network} and Table \ref{DRN}, the computational complexity of the 1st Conv layer in the proposed FL-DRN-NFCE network is 
\begin{align}
C_{f1}=\mathcal{O}(3^2\cdot 2\cdot 32\times(Q+2)\cdot(1+2))
=\mathcal{O}\big(1728(Q+2)\big).
\end{align}
In addition, the calculation method for the complexity of the three residual blocks is similar, as follows
\begin{align}
C_{f2}&=\mathcal{O}
\big((1^2\cdot32\cdot 16+
  3^2\cdot16\cdot 16+
  1^2\cdot16\cdot 64+
  1^2\cdot32\cdot 64)
\times(Q+2)\cdot(1+2)\big)
=\mathcal{O}\big(17664(Q+2)\big),\nonumber\\
C_{f3}&=\mathcal{O}
\big((1^2\cdot64\cdot 32+
  3^2\cdot32\cdot 32+
  1^2\cdot32\cdot 128+
  1^2\cdot64\cdot 128)\times(Q+2)\cdot(1+2)\big)
=\mathcal{O}\big(70656(Q+2)\big),\nonumber\\
C_{f4}&=\mathcal{O}
\big((1^2\cdot128\cdot 64+
  3^2\cdot64\cdot 64+
  1^2\cdot64\cdot 256+
  1^2\cdot128\cdot 256)\times(Q+2)\cdot(1+2)\big)\nonumber\\
&=\mathcal{O}\big(282624(Q+2)\big).
\end{align}
The computational complexity of the Conv layer after three residual blocks is as follows
\begin{align}
C_{f5}=\mathcal{O}\big(1^2\cdot 256\cdot 2\times(Q+2)\cdot(1+2)\big)
=\mathcal{O}\big(1536(Q+2)\big).
\end{align}
The computational complexity of the linear layer of the proposed FL-DRN-NFCE network is as follows
\begin{align}
C_{f6}=\mathcal{O}
\big(\lfloor(Q+1)/2\rfloor\cdot1\cdot2
\times(N+1)\cdot M\cdot 2\big).
\end{align}
Therefore, the total computational complexity of the proposed FL-DRN-NFCE network is
\begin{align}
C_{f}
=C_{f1}+C_{f2}+C_{f3}+C_{f4}+C_{f5}
+C_{f6}
=\mathcal{O}\Big(374208(Q+2)
+4M(N+1)\big(\lfloor (Q+1)/2\rfloor\big)\Big).
\end{align}

\section{Simulation results}

\begin{figure*}[h]
\centering
\includegraphics [width=0.7\textwidth]{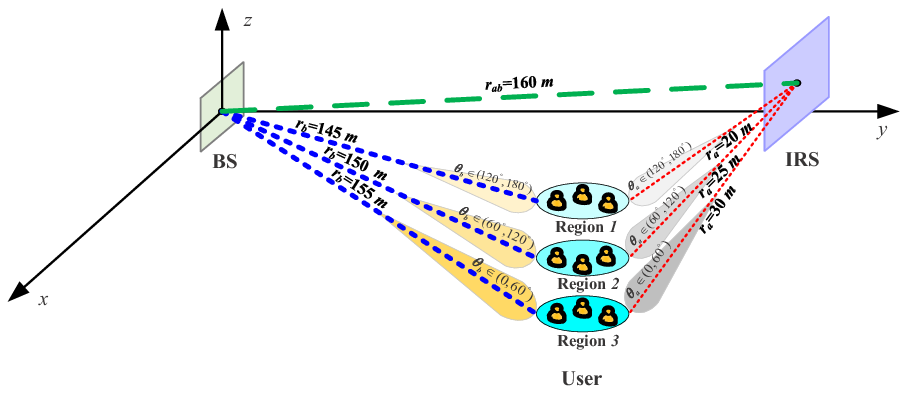}
\caption{Simulation parameters setup for an IRS-aided MU NF communication.}
\label{Simulation-parameters}
\end{figure*}

\begin{table}[h]
\footnotesize
\caption{Simulation parameters setup for the NF scenario.}
\label{Simu-param-tab}
\tabcolsep 25pt 
\begin{tabular*}{\textwidth}{cc}
\toprule
Parameters & Value \\\hline
The azimuth angle of $\mathbf{h}_{t, k}$ in three regions & $\theta_b\in\Big\{(0,\frac{\pi}{3}),
~(\frac{\pi}{3},\frac{2\pi}{3}),~
(\frac{2\pi}{3},\pi)\Big\}$ \\
The elevation angle of $\mathbf{h}_{t, k}$ in three regions & $\phi_b\in(\frac{\pi}{4},\frac{\pi}{2})$ \\
Distance between the user in three regions and BS & $r_b=\{145, 150, 155\}$ m \\
The azimuth angle of $\mathbf{f}_{t, k}$ in three regions & $\theta_a\in\Big\{(0,\frac{\pi}{3}),
~(\frac{\pi}{3},\frac{2\pi}{3}),~
(\frac{2\pi}{3},\pi)\Big\}$ \\
The elevation angle of $\mathbf{f}_{t, k}$ in three regions & $\phi_a\in(\frac{\pi}{4},\frac{\pi}{2})$ \\
Distance between the user in three regions and IRS & $r_a=\{20,25,30\}$ m \\
Distance between BS and IRS & $r_{ab}=r_{ba}=160$ m \\
Carrier frequency &  $f=10$ GHz\\
Wavelength & $\lambda=\frac{c}{f}=\frac{3\times 10^8}{10\times 10^9}=0.03$ m\\
Element spacing of IRS & $d_{xa}=d_{za}=\frac{\lambda}{2}$ \\
Antenna spacing of BS & $d_{xb}=d_{zb}=\frac{\lambda}{2}$\\
Width and height of IRS &  $L_{xa}=N_{xa}\times d_{xa}$ and $L_{za}=N_{za}\times d_{za}$\\
Aperture of IRS & $D_{\text{IRS}}=\sqrt{L_{xa}^2+ L_{za}^2}$ \\    
\bottomrule
\end{tabular*}
\end{table}

In this section, the simulation results are given to substantiate the validity of the proposed FL-DRN-NFCE network.
Firstly, the configuration of IRS assisted MU NF system is shown in Fig.~\ref{Simulation-parameters}, which consists of three regions with three users in each region.
Specifically, $M$ and $N$ are considered to be $M=M_{xb}\times M_{zb}=3\times 41=123$ and $N=N_{xa}\times N_{za}=3\times 41=123$, respectively.
The azimuth angle and elevation angle of the center of BS's UPA relative to the center of IRS's UPA are $\theta_{ba}=\frac{\pi}{4}$ and $\phi_{ba}=\frac{\pi}{4}$.
The other specific parameter configurations are detailed in Table \ref{Simu-param-tab}.
Among them, the azimuths of the $\mathbf{h}_{t, k}$ and $\mathbf{f}_{t, k}$ channels are uniformly distributed in $[0,2\pi)$ and are equally divided into three parts to correspond to three different user regions in the system.
In addition, as depicted in Table \ref{Simu-param-tab}, the NF distances between users in these three regions and BS and IRS are different.
According to \cite{Cui2023Near-field-Mag}, $r_a$ and $r_{ab}$ are consistent with the following equation, then IRS-assisted communication operates in the NF region, 
\begin{align}
\frac{r_{a}\times r_{ab}}{r_{a}+r_{ab}}<R_{\text{RD}}=\frac{2\times D_{\text{IRS}}^2}{
\lambda}.
\end{align}
For comparative analysis, in the FF scenario, the distances between users in the three regions and BS and IRS are set as $r_{b, far}=\{400, 500, 600\}$ m and $r_{a, far}=\{300, 400, 500\}$ m, respectively.
Moreover, the $\mathbf{G}$ channel contains $L_{ab}=15$ NLoS path components, while the NLoS path components of the $\mathbf{h}_{t, k}$ and $\mathbf{f}_{t, k}$ channels in the three regions are $L_b=\{2, 4, 6\}$ and $L_a=\{3, 5, 7\}$, respectively.
Specifically, the distance of the NLoS path components in $\mathbf{h}_{t, k}$ is defined to be generated in range of $(R_{min,b}, R_{max,b})$ for three regions, namely $R_{min,b}=\{20,21,22\}$ and $R_{max,b}=\{100,101,102\}$.
Similarly, for $\mathbf{f}_{t, k}$, we have 
$R_{min,a}=\{15,20,25\}$ and $R_{max,a}=\{20,25,30\}$.
Furthermore, the distance of the NLoS path of $\mathbf{G}$ is generated within the range of $[30, 120]$ and $[25, 100]$.
For the proposed FL-DRN-NFCE network, the required pilot overhead is considered to be set to $Q_1=\frac{M(N+1)}{6}=2542$, while the traditional LS and MMSE algorithms require a pilot overhead of $Q_2=M(N+1)=15252$.

\begin{figure}
\begin{minipage}[h]{0.5\textwidth}
\centering
\includegraphics[width=1\textwidth]
{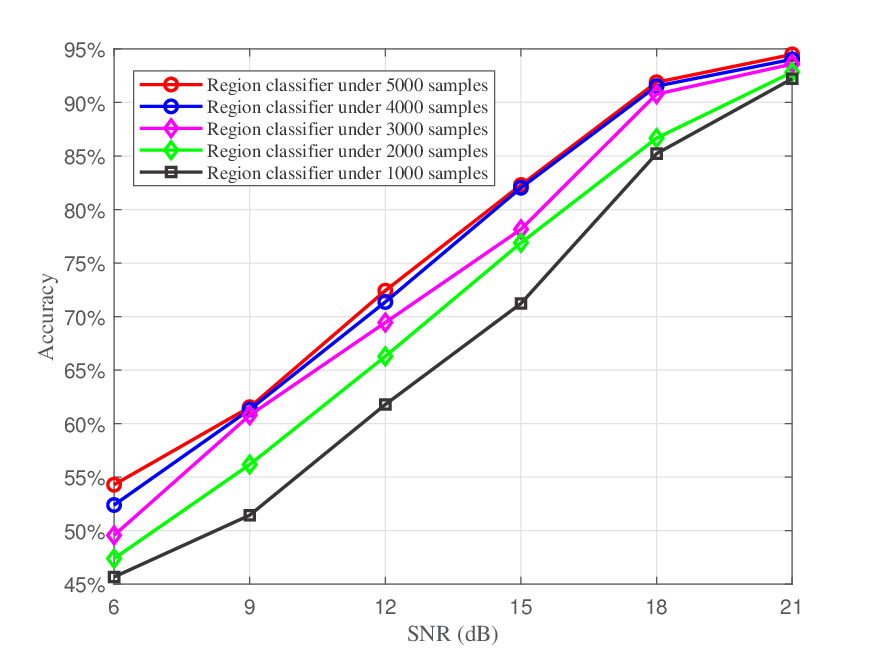}\\
\caption{The accuracy of the proposed RC network.}
\label{accuracy}
\end{minipage}
\begin{minipage}[h]{0.5\textwidth}
\centering
\includegraphics[width=1\textwidth]
{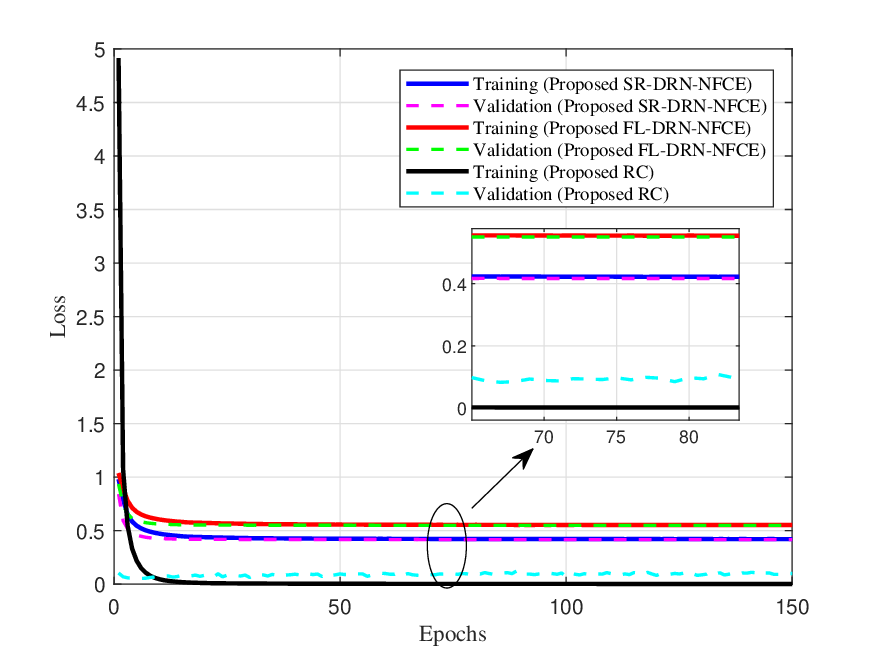}\\
\caption{The training and the validation loss of the proposed RC, DRN-NFCE, and FL-DRN-NFCE network versus the training epochs.}
\label{Loss_epochs}
\end{minipage}
\end{figure}

As shown in Fig.~\ref{Simulation-parameters}, there are a total of 9 users in the 3 regions, and each user collects 5000 training samples. Therefore, for the SR-DRN-NFCE network, $3 \times 5000=15000$ training samples are collected, and for the FL-DRN-NFCE network, $3 \times 3 \times 5000=45000$ training samples are collected.
We allocate 90\% of the samples to construct the training dataset, while the remaining 10\% are reserved for the validation dataset. The AdamW optimization algorithm is employed to update the model's weight parameters, with an initial learning rate of 1e-3. The training process spans a total of 150 epochs, during which the learning rate is halved every 15 epochs until it reaches the minimum threshold of 1e-6. Upon completion of training, the model's performance is evaluated on the validation dataset. Additionally, the SR-DRN-NFCE network is trained with a batch size of 200, whereas the FL-DRN-NFCE network uses a batch size of 256.

Fig.~\ref{accuracy} shows the accuracy of the proposed RC network versus different SNRs.
From Fig.~\ref{accuracy}, it can be observed that as the number of region samples increases, the classification accuracy gradually improves.
When the sample size reaches 4000, continuing to increase the sample size does not improve the accuracy significantly.
Thus, each user is set to collect 5000 samples as a training set.
Moreover, the proposed RC network can reach 95\% accuracy in the high SNR range.

As is clearly evident from Fig.~\ref{Loss_epochs}, the performance of the designed networks has stabilized and converged in terms of loss on both the training and validation sets after 150 epochs, with the losses falling within acceptable ranges. During the training process, to rapidly reduce the networks' losses and effectively mitigate overfitting, we adopted a LR decay strategy. Specifically, at the early stages of training, we initialized the LRs to 1e-3. This relatively high initial value aided the models in quickly capturing the primary features within the data. Following this, we halved the LRs every 15 epochs, progressively decreasing the learning step sizes. The merit of dynamically adjusting the LRs in this fashion is that it not only accelerates the convergence speed of the models, but also significantly boosts their ultimate performances, ensuring stability and accuracy in handling complex tasks.

\begin{figure}
\begin{minipage}[h]{0.5\textwidth}
\centering
\includegraphics[width=1\textwidth]{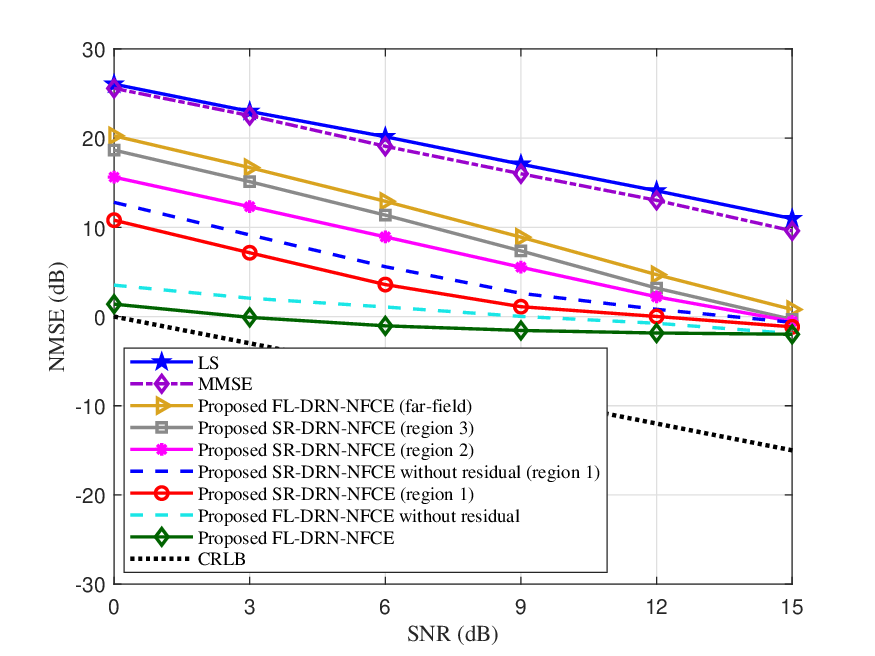}\\
\caption{NMSE performance comparison for the region 1.}
\label{region1}
\end{minipage}
\begin{minipage}[h]{0.5\textwidth}
\centering
\includegraphics[width=1\textwidth]{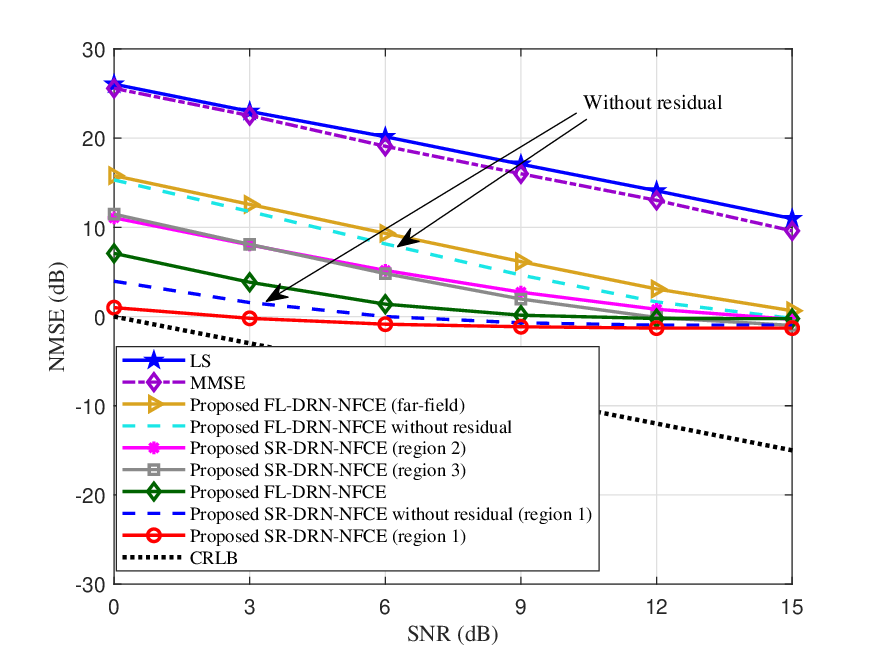}\\
\caption{NMSE performance comparison for the whole region.}
\label{whole-region}
\end{minipage}
\end{figure}

Fig.~\ref{region1} demonstrates the normalized MSE (NMSE) performance comparison of the proposed SR-DRN-NFCE and FL-DRN-NFCE network versus different SNRs on the testing set from region 1.
According to Fig.~\ref{region1}, it can be noticed that the proposed SR-DRN-NFCE network trained only on region 1 achieved the lower MSE on the testing set on region 1.
However, the proposed SR-DRN-NFCE network trained on region 2 and region 3 did not work well on region 1.

Fig.~\ref{whole-region} describes the NMSE performance comparison of the proposed SR-DRN-NFCE and FL-DRN-NFCE network versus different SNRs on the testing set from the whole region.
It can be clearly remarked that the proposed FL-DRN-NFCE network achieved reliable CE in the entire region's test set, which verifies that the training method based on FL can learn different channel characteristics of more users.

From Figs. \ref{region1} and \ref{whole-region}, it can be seen that both the proposed SR-DRN-NFCE and the proposed FL-DRN-NFCE have smaller errors than their corresponding networks without residual connections, which proves the denoising advantage of residual networks.
The LS and MMSE schemes require significant pilot overhead to achieve good estimation accuracy, while the two proposed residual methods can achieve better CE performance while reducing pilot overhead by five-sixth.

\section{Conclusion}

In this paper, the CE performance of an IRS-aided MU NF MIMO system has been investigated.
The LS estimator, MMSE estimator, and CRLB were derived.
Subsequently, to reduce the pilot overhead and improve the CE precision, the SR-DRN-NFCE and FL-DRN-NFCE were proposed.
Compared with traditional methods such as LS and MMSE, the proposed FL-DRN-NFCE network has successfully reduced the pilot overhead by five-sixths.
Simulation results revealed that the proposed FL-DRN-NFCE network presented a lower MSE compared to the scheme without residual connection.
The accuracy of the proposed RC approached 95\% when the number of user samples was 5000.
Finally, the computational complexity of the proposed RC and FL-DRN-NFCE network was delineated.
In the whole region, the proposed methods were in ascending order in terms of MSE: FL-DRN-NFCE, SR-DRN-NFCE, MMSE, and LS. 
In the extreme SNR range, the proposed FL-DRN-NFCE achieves about one-magnitude improvement over SR-DRN-NFCE in terms of MSE performance.
For future works, it is necessary to explore potential solutions to address the performance bottleneck of the proposed method and narrow the gap between the proposed method and CRLB, especially at high SNR levels.

\Acknowledgements{This work was supported in part by the National Natural Science Foundation of China under Grant U22A2002, and  by the Hainan Province Science and Technology Special Fund under Grant ZDYF2024GXJS292; in part by the Scientific Research Fund Project of Hainan University under Grant KYQD(ZR)-21008; in part by the Collaborative Innovation Center of Information Technology, Hainan University, under Grant XTCX2022XXC07; in part by the National Key Research and Development Program of China under Grant 2023YFF0612900; in part by the Innovative Research Project of Postgraduates in Hainan Province under Grant Qhyb2024-17.}






\end{document}